\documentclass[twoside]{article}

\usepackage{blindtext} 

\usepackage{graphicx}
\usepackage[sc]{mathpazo} 
\usepackage[T1]{fontenc} 
\usepackage{microtype} 

\usepackage[english]{babel} 

\usepackage[hmarginratio=1:1, top=32mm, columnsep=20pt]{geometry} 
\usepackage[hang,  small, labelfont=bf, up, textfont=it, up]{caption} 
\usepackage{booktabs} 

\usepackage{lettrine} 
\usepackage{textcomp}
\usepackage{enumitem} 
\setlist[itemize]{noitemsep} 

\usepackage{abstract} 

\usepackage{titlesec} 
\titleformat{\section}[block]{\large\scshape\centering}{\thesection.}{1em}{} 
\titleformat{\subsection}[block]{\large}{\thesubsection.}{1em}{} 

\usepackage{fancyhdr} 
\usepackage{titling} 

\usepackage{hyperref} 

\pretitle{\begin{center}\Huge\bfseries} 
\posttitle{\end{center}} 
\title{Building a COVID-19 Vulnerability Index} 
\author{%
\textsc{Dave DeCaprio}\thanks{Corresponding author: dave.decaprio@closedloop.ai}\\
\normalsize ClosedLoop.ai \\
\and 
\textsc{Joseph Gartner}\thanks{Corresponding author: joseph.gartner@closedloop.ai} \\ 
\normalsize ClosedLoop.ai \\
\and
\textsc{Carol J. McCall, FSA, MPH}\thanks{Corresponding author: carol.mccall@closedloop.ai}\\
\normalsize ClosedLoop.ai \\
\and
\textsc{Thadeus Burgess}\\
\normalsize ClosedLoop.ai \\
\and
\textsc{Kristian Garcia}\\
\normalsize Healthfirst \\
\and
\textsc{Sarthak Kothari}\\
\normalsize ClosedLoop.ai \\
\and
\textsc{Shaayaan Sayed}\\
\normalsize ClosedLoop.ai \\
}
\date{\today} 
\renewcommand{\maketitlehookd}{%
\begin{abstract}
\noindent 
COVID-19 is an acute respiratory disease that has been classified as a pandemic by the World Health Organization. Characterization of this disease is still in its early stages; however, it is known to have high mortality rates, particularly among individuals with preexisting medical conditions. Creating models to identify individuals who are at the greatest risk for severe complications due to COVID-19 will be useful for outreach campaigns to help mitigate the disease's worst effects. While information specific to COVID-19 is limited, a model using complications due to other upper respiratory infections can be used as a proxy to help identify those individuals who are at the greatest risk. We present the results for three models predicting such complications, with each model increasing predictive effectiveness at the expense of ease of implementation. 

\end{abstract}
}

\begin{document}

\maketitle


\section{Introduction}

\subsection{COVID-19 Illness}
Coronaviruses (CoV) are a large family of viruses that cause illnesses ranging from the common cold to more severe diseases such as Middle East respiratory syndrome (MERS-CoV) and severe acute respiratory syndrome (SARS-CoV). CoV are zoonotic, meaning they are transmitted between animals and people. Coronavirus disease 2019 (COVID-19) is caused by a new strain discovered in 2019, severe acute respiratory syndrome coronavirus 2 (SARS-CoV-2), that has not been previously identified in humans\cite{whoCorona}. 

COVID-19 is a highly contagious respiratory infection with common signs that include respiratory symptoms, fever, cough, shortness of breath, and breathing difficulties. In more severe cases, infection can cause pneumonia, severe acute respiratory syndrome, kidney failure, cardiac arrest, and death\cite{sanche}\cite{hawryluk}.  Experts continue to learn more about COVID-19, including its etiology, symptoms, complications, and potential treatments.

\subsection{Flattening the Curve}
On March 11, 2020, the World Health Organization (WHO) declared COVID-19 to be a pandemic\cite{whoPandemic}. Public health and healthcare experts agree that mitigation is required in order to slow the spread of COVID-19 and prevent the collapse of healthcare systems. Health systems in the United States run close to capacity\cite{capacity}, and so every transmission that can be avoided and every case that can be prevented has enormous impact.

\subsection{Identifying Vulnerable People}
The risk of severe complications from COVID-19 is higher for certain vulnerable populations, particularly people who are elderly, frail, or have multiple chronic conditions. The risk of death has been difficult to calculate\cite{detect}, but a small study\cite{study} of people who contracted COVID-19 in Wuhan along with patterns also seen in early reports from the United States suggest that the risk of death increases with age, and is also higher for those who have diabetes, heart disease, blood clotting problems, or have shown signs of sepsis. With an average death rate of 1\%,  the death rate rose to 6\% for people with cancer, high blood pressure, or chronic respiratory disease, 7\% for people with diabetes, and 10\% for people with heart disease. There was also a steep age gradient; the death rate among people age 80 and over was 15\% \cite{study2}.

Identifying who is most vulnerable is not necessarily straightforward. More than 55\% of Medicare beneficiaries meet at least one of the risk criteria listed by the US Centers for Disease Control and Prevention (CDC)\cite{cdcRisk}. People with the same chronic condition don't have the same risk, and many people will have other comorbidities, which compounds their vulnerability.  Simple rules can't reflect these differences or capture complex factors like frailty\cite{frailty} which makes people more vulnerable to severe infections.


\section{Methods}
\subsection{Proxy Outcome}
Since real-world data on COVID-19 cases are not readily available, the C-19 Index was developed using close proxy events. A person's C-19 Index is measured in terms of their near-term risk of severe complications from respiratory infections (e.g. pneumonia, influenza). The most direct proxy event is Acute respiratory distress syndrome (ARDS), identified by ICD-10 diagnosis code J80.  ARDS is extremely rare, with an annual occurence of less than 0.05\% in Medicare members.  To create a viable machine learning model, the outcome was broadened to include 4 closely related categories of respiratory diagnoses from the Clinical Classifications Software Refined (CCSR)\cite{ccsr} classification system.  Patient's were considered to have the proxy event if they had any of the following diagnosis codes in any position on a medical claim associated with a hospital inpatient visit or observation stay:
\begin{itemize}
   \item ICD-10-CM J80 - Acute respiratory distress syndrome
   \item RSP002 - Pneumonia (except that caused by tuberculosis)
   \item RSP003 - Influenza
   \item RSP005 - Acute bronchitis
   \item RSP006 - Other specified upper respiratory infections
\end{itemize}
Machine learning models were created that predict the likelihood that a patient will have an inpatient hospital stay due to one of the above conditions in the next 3 months.  While claims-based machine learning models typically focus on longer term predictions such as outcomes with the next 6 months or one year, this model is expected to be used for immediate targeting decisions around COVID-19, and so a shorter window was deemed appropriate.

\subsection{Datasets}
These models were trained and initially tested using historical medical claims data.  Two different data sets were obtained, representing different segments of the population.  The first was the Center for Medicare \& Medicaid Services (CMS) Limited Data Set (LDS) \cite{cmslds} for 2015 \& 2016.  The LDS contains beneficiary level health information for 5\% of the Medicare population, and is available to the public subject to a Data Use Agreement.  The second was medical claims data for 2.5 million beneficiaries obtained from Healthfirst, which provides health insurance for New Yorkers.  These two data sets represented different demographics within the US population.  The LDS data contains only Medicare beneficiaries, who are predominantly over the age of 65 or disabled.  The Healthfirst data set is primarily a Medicaid population, which includes more healthy adults.

In order to build a predictive model appropriate for the overall US population, these two data sets were combined into a single population that has a demographic profile consistent with the overall US population.  Cohorts were individually created for each data set and then the resulting cohorts were combined to create the final training and test sets for the models.

\subsubsection{CMS Data Preparation}
The CMS data cohort was created by identifying all living members above the age of 65 on 9/30/2016.  This particular date was chosen because it was 3 months from the end of the data set and was therefore the latest prediction date possible in the data set.  Using the latest date minimized the use of ICD-9 data in the prediction histories since this data spanned the transition from International Classification of Diseases version 9 to version 10 (ICD-10) on October 1, 2015.  Members under 65 were excluded because the second population we had was a better representation of the under 65 population.  Only fee-for-service members were included because medical claims histories for other members are not reliably complete. We then excluded all members who had less than 6 months of continuous eligibility prior to 9/30/2016. We also excluded members who lost coverage within 3 months after 9/30/2016, except for those members who lost coverage due to death.  Because these members lost coverage, we cannot be confident that a negative label s  Table \ref{tab:pop-sel-cms} below summarizes the population selection.

\begin{table}[tbp]
\begin{tabular}{ll}
\textbf{Population Size} & \textbf{Selection Criteria} \\
    3,114,713 & Total members in the CMS dataset\\
    1,867,879 & Fee-for-service members with 6 months of continuous coverage prior to 9/30/2016\\
    1,511,950 & 65 years old or older\\
    1,506,659 & Exclude members who died before 9/30/2016\\
    1,500,700 & Exclude members who lose coverage before 12/31/2016 not due to death.\\
\end{tabular}
\caption{Cohort selection for the CMS population}
\label{tab:pop-sel-cms}
\end{table}

\subsubsection{Healthfirst Data Preparation}
The Healthfirst cohort contained a longer history, from 2017 to the present.  For this data set members were evaluated at multiple points in time rather instead of having a single prediction date.  Each person was evaluated using each month of eligibility as a prediction date.  This approach was not used with the CMS data because that data didn't cover a long enough time period to allow for multiple months.  Members had to have 3 months of eligibility after and had to be at least 18 years old on the prediction date.  Table \ref{tab:pop-sel-hf} below summarizes the population selection.

\begin{table}[tbp]
\begin{tabular}{ll}
\textbf{PopulationSize} & \textbf{Selection Criteria} \\
3,008,781 & Total members in the Healthfirst dataset\\
30,898,086 & Total member-months of eligibility \\
29,388,003 & 3 months of eligibility after the prediction date\\
19,470,511 & 18 years or older on the prediction date  \\
\end{tabular}
\caption{Cohort selection for the Healthfirst population}
\label{tab:pop-sel-hf}
\end{table}

\subsubsection{Combined Population}
Each data set was split by person 80\%/20\% into train and test sets.  In the Healthfirst data set, doing the split by person ensures that all prediction dates for a given individual were in the same set, eliminating the possibility of data leakage.  The data set for training was taken by taking all of the positive examples from each data set along with a sampling of the negative examples in the training set.  The sampling was designed to meet two criteria.  First, the percentage of the adult population 65 or older was set to match that of the US population (21\% of adults).  Second, the difference in prevalence between those under and over 65 from the Healthfirst data set was maintained in the overall data set.   This difference was 3.9X.

The combined test population was created by unioning the full test set from the CMS data and a random 20\% sample of the Healthfirst test set.  The full test population contained 1,621,149 training examples with a prevalence of 3.86\%.  The test population had 761,898 examples with a prevalence of 0.36

\subsection{Models}
In order to encourage wide adoption and rapid implementation of the predictive models, we created three separate models which represent different tradeoffs between accuracy and ease of implementation.  All three models were trained and tested on the same data set.  The models were built using the ClosedLoop platform on the AWS cloud.  The models are summarized in Table \ref{tab:models} below.

\begin{table}[tbp]
\begin{tabular}{lllll}
\textbf{Model} & \textbf{Algorithm} & \textbf{Deployment} & \textbf{Features} & \textbf{SLA @ 5\%} \\
Survey & Logistic Regression & Online survey & 14 & 49.8\% \\
Open Source & XGBoost & Open Source &  559 & 53.8\% \\
Full & XGBoost & Hosted & 892 & 54.1\% \\
\end{tabular}
\caption{Different models created and their tradeoffs in terms of accuracy versus ease of implementation.  The full model has the highest accuracy while the survey model has a very constrained feature set that is easy to implement.  Accuracy is measured using Sensitivity and Low Alert Rates (SLA).}
\label{tab:models}
\end{table}

The "Survey" model is the simplest and uses only a small number of features that were designed to be able to be generated from a simple health risk assessment questionnaire.  This model requires no technical implementation, and we have made it available through a web-based survey\cite{c19survey}.  The "Open source" model uses approximately 600 features derived from medical claims diagnosis and utilization data.  We have made this model available on GitHub\cite{github}.  Finally, the "Full" model was created that uses an extensive feature set derived generated from medical claims data along with linked geographical and social determinants of health data. This model is being made freely available to healthcare organization.  Information about accessing the platform can be found at https://cv19index.com.

\subsubsection{Survey Model}
The first model is aimed at using a simple health history survey to enrich the high-level recommendations from the CDC website\cite{cdcGuide} for identifying those individuals who are at risk. The CDC identifies risk factors as:
\begin{itemize}
   \item Older adults
   \item Individuals with heart disease
   \item Individuals with diabetes
   \item Individuals with lung disease
\end{itemize}

The purpose of the survey to to let an individual know their risk relative to the general population with more detail than is available through the CDC.  The way this is achieved is by mapping questions related to an individuals medical history into diagnosis code categories from the CCSR.  We also included age and gender as well as prior year hospital inpatient or emergency room (ER) visits. In addition to the conditions coming from the recommendations of the CDC, we included features that our other modeling efforts surfaced as important.   The mapping between the survey questions and CCSR codes is described in Table \ref{tab:cdc-css}.  To turn this into a model, we extract ICD-10 diagnosis codes from the claims in the year before the prediction date and aggregate them using the CCSR categories. We create indicator features for the presence of any code in the CCSR category.  A logistic regression model is then trained on the available claims data.  A person's percentile risk score is based on risk relative to the other values in the training distribution.   In addition to the CCSR codes, Table \ref{tab:cdc-css} includes the coefficients associated with these features in the logistic regression model.

\begin{table}[]
\begin{tabular}{lll}
\textbf{Feature name} & \textbf{Coefficient} & \textbf{CCSR Code}\\
Intercept & -6.74 & n/a \\
Age & 0.041 & n/a \\
Gender Male & 0.171 & n/a \\
Prior Admissions & 0.682 & n/a \\
Prior ER Visits & 0.413 & n/a \\
Chronic obstructive pulmonary disease (COPD) or & 1.167 & CCSR:RSP008,CCSR:END012 \\
emphysema, cystic fibrosis, or chronic bronchitis & & \\
Asthma & 1.393 & CCSR:RSP009 \\
Obesity & 0.935 & CCSR:END009 \\
Diabetes (other than when you were pregnant) & 0.096 & CCSR:END002,CCSR:END003, \\
 & & CCSR:END004,CCSR:END005 \\
Hypertension (also called high blood pressure) & 0.832 & CCSR:CIR007,CCSR:CIR008 \\
Congestive Heart Failure & 0.982 & CCSR:CIR019 \\
Heart attack (also called myocardial infarction) & 0.159 & CCSR:CIR009,CCSR:CIR010 \\
Rheumatic heart disease & 0.788 & CCSR:CIR001,CCSR:CIR002,CCSR:CIR011, \\
 & & CCSR:CIR014,CCSR:CIR015 \\
Stroke & 0.285 & CCSR:CIR020,CCSR:CIR021 \\
Sickle cell anemia/HIV infection/Transplant & 2.582 & CCSR:BLD005,CCSR:INF006,CCSR:FAC023 \\
Chronic kidney disease & 0.966 & CCSR:GEN003 \\
Hemodialysis & 1.369 & CCSR:GEN002 \\
Liver disease & 0.055 & CCSR:DIG019 \\
Pneumonia, acute bronchitis, influenza or other & 0.696 & CCSR:RSP002,CCSR:RSP003, \\
 & & CCSR:RSP005,CCSR:RSP006 \\
acute respiratory infection &  &  \\
Cancer & 1.091 & CCSR:NEO \\
Neurocognitive conditions & 0.294 & CCSR:NVS011,CCSR:CIR022,CCSR:CIR025 \\
Pregnancy & 0.789 & CCSR:PRG \\
COPD x Age & -0.002 & n/a \\
Asthma x Age & -0.015 & n/a \\
Obesity x Age & -0.004 & n/a \\
Diabetes x Age & 0.000 & n/a \\
Hypertension x Age & 0.005 & n/a \\
Congestive heart failure x Age & -0.007 & n/a \\
Myocardial infarction x Age & 0.003 & n/a \\
Rheumatic heart disease x Age & -0.008 & n/a \\
Stroke x Age & -0.003 & n/a \\
Sickle cell/HIV/Transplate x Age & -0.028 & n/a \\
Chronic kidney disease x Age & -0.008 & n/a \\
Hemodialysis x Age & -0.018 & n/a \\
Liver disease x Age & 0.001 & n/a \\
Pneumonia, acute bronchitis, influenza or other & -0.005 & n/a \\
acute respiratory infection x Age &  &  \\
Cancer x Age & -0.009 & n/a \\
Neurocognitive conditions x Age & 0.004 & n/a \\
Pregnancy, childbirth and the puerperium x Age & -0.003 & n/a \\
\end{tabular}
\caption{Features used associated with risk factors identified by CDC \\ and their corresponding CCSR codes}
\label{tab:cdc-css}
\end{table}

\subsubsection{Open Source Model}
The Open Source model uses gradient boosted trees. Gradient boosted trees are a machine learning method that use an ensemble of simple models to create highly accurate predictions\cite{boosting}.  The resulting models demonstrate higher accuracy. A drawback to these models is that they are significantly more complex; consequently, "by hand" implementations of such models are impractical. A nice feature of gradient boosted trees is that they are fairly robust against learning features that are eccentricities of the training data, but do not extend well to future data. As such, we allow full diagnosis histories to be leveraged within our simpler XGBoost model. In this approach, every category in the full CCSR is converted into an indicator feature, resulting in 559 features. A 3-month delay was imposed on the claims data, so that claims within the most recent 3 months before the prediction date were not used to make the predictions. This 3-month delay simulates the delay in claims processing that usually occurs in practical settings and enables the model to be used with current claims data.  This delay was not imposed in the survey model since questionnaires would generally use current data.  The Github repository for the open source model contains scripts that automatically prepare the features for this model from simple CSV files containing claims data.

\subsubsection{Full Model}
We additionally built a model within the ClosedLoop platform. The ClosedLoop platform is a software system designed to enable rapid creation of machine learning models utilizing healthcare data. The full details of the platform are outside the bounds of this paper; however, using the platform allows us to leverage engineered features coming from peer-reviewed studies. Examples are social determinants of health and the Charlson Comorbidity Index\cite{charlson}. The computation of these features from claims data is often complex or involves the linking of additional data.  These are operations handled by the CLosedLoop platform that were not easy to extract into an open source format, but do provide improved model accuracy.

\section{Results and Model Interpretation}
\begin{figure}
\centering
\includegraphics[width=0.75\textwidth]{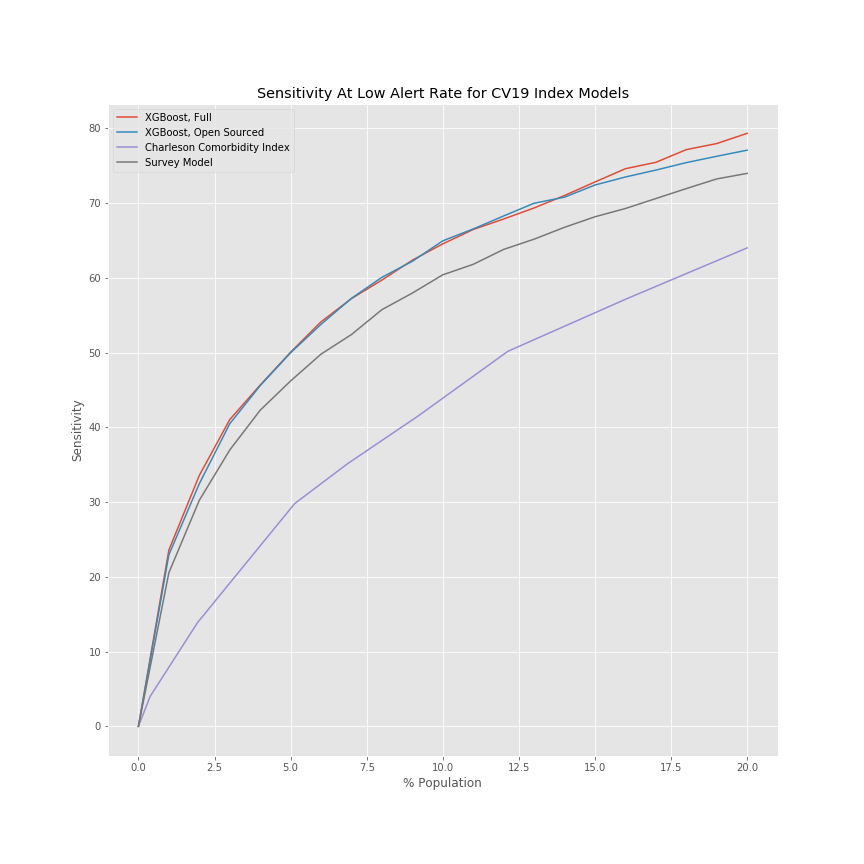}
\caption{Sensitivity of each model at alert rates up to 20\%.}
\label{fig:sla}
\end{figure}

We quantify the performance of the C-19 Index using metrics that are relevant for it's intended use. The standard receiver operating characteristic curves show that all 3 models have identical areas under the curve, at 0.87.  However, if we consider the intended use of the model, which is to identify highly vulnerable populations for additional targeting, a more appropriate metric is sensitivity at low alerts rates.  This metric measures the sensitivity of the model when looking only at some small percentage of the population.  This sensitivity if plotted for alert rates up to 20\% of the population in Figure \ref{fig:sla}. Additionally, the metrics quantifying the effectiveness of our models are presented in Table \ref{tab:model}.  To provide a baseline for comparison, the models are compared to the Charleson Comorbidity Index (CCI)\cite{charlson}.  The CCI is a typical risk scoring algorithm based on claims data.  In this case, we see that models explicitly built for the proxy outcome are more accurate than the CCI.

\begin{table}[]
\begin{tabular}{llll}
\textbf{Model} & \textbf{ROC AUC} & \textbf{Sensitivity at 5\% Alert Rate}\\
Survey & .86 & .498\\
Open Source & .87 & .538 \\
Full & .87 & .541 \\
Charlson Comorbidity & .79 & .291 \\
\end{tabular}
\caption{Measures of effectiveness for the three models}
\label{tab:model}
\end{table}

\subsection{Validation}
The model has subsequently been validated by evaluating its results against approximately 14,000 hospital admissions for known COVID-19 cases in New York City from 2/1/2020 until mid-May 2020.  These admissions were We validated the Vulnerability Index by comparing the mortality rate for these admissions with their Vulnerability Index.  All cases were insured by Healhfirst, and each patient's prior claims data was used to compute their Vulnerability Index.

Members who were admitted and survived had an average score of 2.4\% and members who were admitted and passed away had an average score of 3.3\% (a 38\% relative difference).  The ROC AUC was 0.68.  This is lower than the test ROC, but that drop  is expected since this test set consisted only of patients who had already been admitted for COVID-19, presumably removing many of the low vulnerability patients and increasing the difficulty of the prediction problem.

\begin{table}[]
\begin{tabular}{lllll}
\textbf{Decile} & \textbf{Medicaid Mortality \%} & \textbf{Medicaid Lift} & \textbf{Medicare Mortality \%}  & \textbf{Medicare Lift}\\
Top 10\% & 6.48 & 1.30 & 5.56 & 2.99 \\
Top 20\% & 6.43 & 1.29 & 3.70 & 2.00 \\
Top 30\% & 6.39 & 1.29 & 3.09 & 1.66 \\
Top 40\% & 6.39 & 1.28 & 2.78 & 1.50 \\
Top 50\% & 6.21 & 1.25 & 2.59 & 1.40 \\
Top 60\% & 5.80 & 1.17 & 3.10 & 1.67 \\
Top 70\% & 5.51 & 1.11 & 2.65 & 1.43 \\
Top 80\% & 5.25 & 1.06 & 2.32 & 1.25 \\
Top 90\% & 5.11 & 1.03 & 2.06 & 1.11 \\
Full & 4.97 & - & 1.86 & - \\
\end{tabular}
\caption{Validation of the Vulnerability Index using COVID-19 admissions}
\label{tab:valid}
\end{table}

The 14,000 cases were divided into a adult Medicaid population and a Medicare population.  Table \ref{tab:valid} sorts each population by their vulnerability index and shows the detah rate of each decile along with the lift relative to the overall death rate.  

\subsection{Accessing Models}
In an effort to make these models as broadly available as possible we have provided several different avenues for the models to be used, each optimized for a different user base.  The logistic regression model powers a publicly available web-based survey at http://c19survey.closedloop.ai.  The open source model is available through github at https://github.com/closedloop-ai/cv19index. This model is written in the Python programming language. We have included synthetic data for testing an wrapper code that converts tabular medical claims data to the input format specific for our models. We encourage the healthcare data science community to fork the repository and adapt it to their own purposes. We encourage collaboration from the open-source community, and pull requests will be considered for inclusion in the main branch of the package.  Finally, for those wishing to take advantage of the full model, we are providing access to the COVID-19 model hosted on the ClosedLoop platform free of charge. Please visit https://closedloop.ai/cv19index for instructions on how to gain access.

\subsection{Limitations}
The approach taken in this paper has several limitations.  Most notably, no actual COVID-19 cases were used in the training of the model.  The usefulness of the model in predicting COVID-19 vulnerability is entirely dependent on the actual occurrence of COVID-19 matching the proxy outcome.  While the logic behind these decisions is defensible, they are not currently backed up with actual data.  As COVID-19 case data becomes available, we expect to validate the proxy outcome and determine if it is in fact appropriate.  Eventually, enough data will be available to build models on COVID-19 Vulnerability itself without having to use a proxy.

Anther major limitation of these models is their reliance on claims data, which does not contain enough clinical detail, such as lab values.  For this reason, we do not recommend using the C-19 Vulnerability Index in inpatient settings, where more detailed data is likely available.  The C-19 Index is most useful in a population health context where the only data available is claims data.

Finally, bassed on medical guidance the authors have decided exclude pediatric populations from the training and test sets for the C-19 Index.  At the time of development, there was so little information available on COVID-19 that we could not confidently assert that the proxy endpoint we were using was appropriate for those under 18 years of age.

\section{Conclusions}
This pandemic has already claimed tens of thousands of lives as of this writing, and sadly this number is sure to grow. As healthcare resources are constrained by the same scarcity constraints that affect us all, it is important to empower intervention policy with the best information possible. We have provided several implementations of the C-19 Index and means of access for those individuals with varying levels of technical expertise. It is our hope that by providing this tool quickly to the healthcare data science community, widespread adoption will lead to more effective intervention strategies and, ultimately, help to curtail the worst effects of this pandemic.

\section{Addendum}
Due to the early release of the first versions of the C19 Index, several organizations were able to quickly apply the model to their populations.  The authors have been in contact with several organizations that have used the vulnerability index to prioritize proactive outreach towards the most vulnerable members.  In many cases, these organizations had existing care management teams that were able to rapidly shift to COVID-19 by applying a different prioritization and focus to their interventions.  In other cases, new phone or text messaging campaigns were developed for COVID-19 that used the C-19 Index as a prioritization mechanism.  As the response to the pandemic continues to develop, we will continue to update the C-19 Index and provide more information on its usage.

\section{Ethics}
All relevant ethical guidelines have been followed in this study.  For the CMS LDS data, we submitted a research proposal as part of our data use agreement that discussed our plans to build predictive models using this data.  For the Healthfirst data, which did contain identifiable PHI, the data was covered under a Business Associate Agreement between ClosedLoop.ai and Healthfirst and the data was stored in ClosedLoop's HIPAA comliant cloud storage.  The project was part of an ongoing Quality Improvement effort at Healthfirst and so did not require preapproval by an Institutional Review Board.  The study protocol and this manuscript were reviewed and approved by Healthfirst's compliance team.

\section{Acknowledgements}
We would like to thank Healthfirst for their collaboration on this work and for allowing us to use insights from their dat ain generation of the model.  We would also like to thank Amazon Web Services for sponsoring this work with AWS platform credits.


‌
\section{Appendix: Full Feature List}
We include the list of features used in the Full model.  This was a feature set that could be implemented across the 2 data sets and utilizes just medical claims data.  The ClosedLoop platform is capable of incorporating a wide variety of other health data, including pharmacy claims, electronic health records, patient reported data, and linked census data.  That additional data was not used in this model due ot the constraints around our usage of the CMS data. The majority of features are binary variables indicating if a patient has had one type of medical event 15 moths prior to the date of prediction, excluding the 3 most recent months.
\begin{table}
\begin{tabular}{ll}
Gender
Age
Admission Count
ED Visit Count
Office Visit Count
Medical Eligibility Months
\# Distinct DME Categories (12M)
\# Days Since Last Hospital Discharge
Use of Preventative Services
Procedure History
DME
\# ICU Stays (12M)
Frailty Indicator
Discharge Disposition
Diagnosis History
Monthly Medical Cost
\# of Admissions (12M)
Inpatient Days
\# of Observation Stays
\# of ER Visits (12M)
\# of ER Visits (6M)
Admission Cause
Charlson Comorbidity Index (CCI)
Evidence of tobacco smoking
Evidence of smokeless tobacco
Evidence of tobacco non-use
Falls risk assessment performed
Abnormal BMI
Normal BMI
LDL
LVEF
Systolic Blood Pressure
Diastolic Blood Pressure
\# Distinct CCSR Body Systems
\# Distinct CCSR Diagnosis Categories (12M)
PCS Procedure History
Prior Respiratory Infections
Prior Hospital Acquired Infections
Diagnosis History High-Level
\end{tabular}
\caption{Full feature list available within platform}
\end{table}

\begin{table}
\begin{tabular}{ll}
Evidence of smokeless tobacco\\
Evidence of tobacco non-use\\
Falls risk assessment performed\\
Abnormal BMI\\
Normal BMI\\
LDL\\
LVEF\\
Systolic Blood Pressure\\
Diastolic Blood Pressure\\
Continuity of Care Index (12M)\\
PQI Any\\
PQI Diabetes\\
PQI 5\\
PQI 7\\
PQI 8\\
PQI 11\\
PQI 12\\
\# Distinct CCSR Body Systems\\
\# Distinct CCSR Diagnosis Categories (12M)\\
PCS Procedure History\\
Prior Respiratory Infections\\
Prior Hospital Acquired Infections\\
Diagnosis History High-Level
\end{tabular}
\caption{Full feature list available within platform (continued)}
\end{table}
\end{document}